\pdfoutput=1
\documentclass[a4paper,11pt,twocolumn]{article} 
\usepackage[utf8x]{inputenc} 
\usepackage[english]{babel}
\usepackage{amsmath,amsthm,amsfonts,amssymb,mathrsfs, bbm}
\usepackage{authblk}
\usepackage{lineno}
\usepackage[dvipsnames]{xcolor}
\usepackage{hyperref}
\usepackage{fourier}

\definecolor{BeauBlue}{rgb}{0, 0.2, .9}
\definecolor{BeauOrange}{rgb}{.8, .1, 0}
\usepackage{hyperref}
\hypersetup{
	colorlinks = true,
	linkcolor = BeauBlue,
	urlcolor = cyan,
	citecolor = BeauOrange,
}

\numberwithin{equation}{section}

\DeclareMathOperator{\Tr}{Tr} 

\title{Response functions of many-body condensed matter systems}
\author[1]{Marcello Porta}
\author[2]{Vieri Mastropietro}
\author[3]{Alessandro Giuliani}
\affil[1]{Mathematics Area, SISSA, Via Bonomea 265, 34136 Trieste, Italy}
\affil[2]{Mathematics Department, University of Milan, Via C. Saldini 50, 20133 Milan, Italy}
\affil[3]{Mathematics and Physics Department, Roma 3 University, Largo S. L. Murialdo 1, 00146 Roma, Italy}

\date{\today}

\begin{document}

\maketitle

\begin{abstract}
We discuss rigorous results about the response functions of gapped and gapless interacting fermionic lattice models, relevant for the study of transport in condensed matter physics.
\end{abstract}

\tableofcontents

\section{Introduction}

A large part of modern condensed matter physics is devoted to the study of the transport properties of many-body quantum systems. In many physical applications, the collective behavior of complex systems can only be understood according to the laws of quantum mechanics. Despite the complexity of the microscopic structure of the material, in some cases the response properties of the system appear to be universal, that is they depend only on fundamental constants of nature. The paradigmatic example is the integer quantum Hall effect, where the quantization of the transverse conductivity of a class of ultrathin materials has a topological interpretation. Other examples include the minimal optical conductivity in graphene and the magneto-electric effect in Weyl semimetals, whose universal nature can be understood via the mechanism of anomaly non-renormalization, a key ingredient in the quantization of gauge theories.

Because of the extremely accurate experimental precision in the measurement of such transport phenomena, it is important to develop a theoretical understanding that does not rely on uncontrolled approximations or ad hoc regularizations. In this view, mathematical physics plays an important role in the fundamental understanding of the behavior of complex systems from first principles. As a matter of fact, the study of transport in condensed matter physics provided a strong motivation for many works in mathematical physics in recent years, and it stimulated cross-fertilization between various areas of mathematics and physics. As a concrete example, our precise understanding of the quantum Hall effect, and nowadays of the field of topological phases of matter, has been made possible by insights coming from various branches of mathematics, ranging from the theory of topological invariants to functional analysis and probability. 

In this article we will give an overview of rigorous results about the transport properties of many-body condensed matter systems. The framework will be the one of quantum statistical mechanics of interacting lattice models of fermionic type. We will give an overview of some of the recent progress, which has been obtained via the combination of several tools in the analysis of many-body quantum systems, such as Lieb-Robinson bounds, cluster expansion, renormalization group, adiabatic theory and Ward identities.

The article is organized as follows. In Section \ref{sec:intro} we will define the class of lattice models we will consider, in Fock space. In Section \ref{sec:dyn} we will introduce the time-evolution of equilibrium Gibbs states under external perturbations, we will define linear response, and we will briefly review rigorous results about the validity of linear response. Then, in the rest of the article we will discuss rigorous results about the computation of response functions for many-body systems. In Section \ref{sec:QHE} we will discuss gapped many-body systems and the quantization of the Hall conductivity; in Section \ref{sec:semi} we will focus on interacting gapless systems of semimetallic type; while in Section \ref{sec:1d} we will discuss metallic transport, for interacting one-dimensional systems and for the edge currents of two-dimensional many-body quantum Hall systems.

\section{Interacting lattice models}\label{sec:intro}

We shall describe interacting fermionic particles on a $d$-dimensional lattice, as models for electrons on crystalline solids in the tight-binding regime. An interesting question, not discussed in this article, is to justify the tight-binding approximation from the continuum Schr\"odinger equation, in the presence of periodic potentials describing the location of the ions forming the crystal. We refer the reader to {\it e.g.} \cite{Ca, DiSj, HS, Ne, PST} and references therein, for the tight-binding reduction in a semiclassical regime, or to the more recent works \cite{FW, FLW}, for the discussion of graphene-like structures.

We shall describe such lattice models in the language of statistical mechanics, in the grand-canonical ensemble. Let $\Gamma_{L}$ be a $d$-dimensional lattice of side $L$ with periodic boundary condition, $\Gamma_{L} = \mathbb{Z}^{d} / L \mathbb{Z}^{d}$. We will suppose that each site is decorated by $M$ colors, that might take into account internal degrees of freedom such as the spin or the sublattice label.  We shall collect together space labels and color labels in the decorated lattice $\Lambda_{L} = \Gamma_{L} \times \{1, \ldots, M\}$. We shall use the notation ${\bf x} = (x, \sigma)$ for points on $\Lambda_{L}$, with $x\in \Gamma_{L}$ the space label and $\sigma \in \{1, \ldots, M\}$ the color label.

The single-particle Hilbert space for one quantum particle on $\Lambda_{L}$ is given by $\frak{h} =\ell^{2}(\Lambda_{L})$, which we can view as $\mathbb{C}^{M L^{d}}$. The dynamics is generated by a self-adjoint Hamiltonian $H$ on $\frak{h}$, via the Schr\"odinger equation,
\begin{equation}\label{eq:S}
i\partial_{t} \psi_{t} = H\psi_{t}\;,\qquad \psi_{0} = \psi \in \mathbb{C}^{M L^{d}}\;.
\end{equation}
where we set $\hbar = 1$.

Eq. (\ref{eq:S}) describes the dynamics of one particle on $\Lambda_{L}$. Instead, we will be interested in describing many interacting fermionic particles on $\Lambda_{L}$. We will work in a grand-canonical setting, in which the number of particles is not fixed. To this end, we introduce the fermionic Fock space as $\mathcal{F} = \bigoplus_{n} \frak{h}^{\wedge n}$. Notice that, being $L$ and $M$ finite, the fermionic Fock space $\mathcal{F}$ is finite dimensional. Vectors in $\mathcal{F}$ are (finite) sequences of complex-valued antisymmetric functions, $\psi = (\psi^{(0)}, \psi^{(1)}, \ldots, \psi^{(n)}, \ldots)$, labelled by the particle number $n$. Let us introduce the fermionic creation and annihilation operators $a^{*}_{{\bf x}}$ and $a_{{\bf x}}$, for ${\bf x}\in \Lambda_{L}$, satisfying the canonical anticommutation relations \cite{BR1, BR2}:
\begin{equation}
\{ a^{*}_{{\bf x}}, a_{{\bf y}} \} = \delta_{{\bf x}, {\bf y}}\;,\qquad \{ a_{{\bf x}}, a_{{\bf y}} \} = \{ a_{{\bf x}}, a_{{\bf y}} \} = 0\;.
\end{equation}
Given the vacuum vector $\Omega = (1, 0, \ldots, 0)$, the fermionic Fock space can be generated by iterative application of the creation operators on the vacuum. Also, operators on $\mathcal{F}$ can be represented as polynomials in the creation and annihilation operators. A simple example is the number operator $\mathcal{N} = \sum_{{\bf x} \in \Lambda_{L}} a^{*}_{{\bf x}} a_{{\bf x}}$, which counts the number of particles in a given sector of the Fock space: $(\mathcal{N} \psi)^{(n)} = n \psi^{(n)}$.

The dynamics in Fock space is generated by a self-adjoint Hamiltonian $\mathcal{H}$ on $\mathcal{F}$. Given an operator $\mathcal{O}$, its Heisenberg evolution generated by $\mathcal{H}$ will be denoted by $\tau_{t}(\mathcal{O})$:
\begin{equation}
\tau_{t}(\mathcal{O}) = e^{i\mathcal{H}t} \mathcal{O} e^{-i\mathcal{H}t}\qquad \text{$t\in \mathbb{R}$}\;.
\end{equation}
A typical form of Hamiltonian, that includes many physically relevant cases, is
\begin{equation}\label{eq:H0}
\mathcal{H} = \sum_{X\subseteq \Lambda_{L}} \Phi_{X}\;,
\end{equation}
with $\Phi_{X} = \Phi_{X}^{*}$ given by a polynomial in the fermionic creation and annihilation operators labelled by ${\bf x} \in X$, vanishing for $\text{diam}(X) > R$ for some $R>0$. As $L$ varies, Eq. (\ref{eq:H0}) actually defines a sequence of Hamiltonians labelled by $L$; if $R$ is independent of $L$, we shall say that $\mathcal{H}$ is finite-ranged. An important example of such Hamiltonians is provided by many-body perturbations of non-interacting models,
\begin{equation}\label{eq:H}
\begin{split}
\mathcal{H} &= \mathcal{H}_{0} + \lambda \mathcal{V} \\
\mathcal{H}_{0} &= \sum_{{\bf x}, {\bf y} \in \Lambda_{L}} a^{*}_{{\bf x}} H_{0}({\bf x}; {\bf y}) a_{{\bf y}}\\
\mathcal{V} &= \lambda \sum_{{\bf x}, {\bf y} \in \Lambda_{L}} v({\bf x}; {\bf y}) a^{*}_{{\bf x}} a^{*}_{{\bf y}} a_{{\bf y}} a_{{\bf x}}\;,
\end{split}
\end{equation}
with $H_{0}, v$ finite-ranged. The quadratic operator $\mathcal{H}_{0}$ is the second quantization of the single-particle Hamiltonian $H_{0}$, while the quartic operator $\mathcal{V}$ is the second quantization of the many-body interaction, specified by the two-body potential $v$ and coupling strength $\lambda$.

The Gibbs state of the system, describing its equilibrium state at inverse temperature $\beta$ and chemical potential $\mu$, is denoted by $\langle \cdot \rangle_{\beta, \mu, L}$, and it is defined as:
\begin{equation}\label{eq:O}
\langle \mathcal{O} \rangle_{\beta, \mu, L} = \frac{\Tr_{\mathcal{F}} \mathcal{O} e^{-\beta (\mathcal{H} - \mu \mathcal{N})}}{ \Tr_{\mathcal{F}} e^{-\beta( \mathcal{H} - \mu \mathcal{N} )} }\;.
\end{equation}
Eq. (\ref{eq:O}) defines the statistical average of the observable associated with the operator $\mathcal{O}$. Being $\mathcal{F}$ finite dimensional, the expression (\ref{eq:O}) is obviously well-defined. However, at this level of generality it is typically extremely hard to extract information from the Gibbs state. For instance, given two observables $\mathcal{O}_{X}$, $\mathcal{O}_{Y}$, a typical question is about decay of their connected correlation function at equilibrium:
\begin{equation}\label{eq:OXOY}
\langle \mathcal{O}_{X}; \mathcal{O}_{Y} \rangle_{\beta, \mu, L} = \langle \mathcal{O}_{X} \mathcal{O}_{Y} \rangle_{\beta, \mu, L} - \langle \mathcal{O}_{X} \rangle_{\beta, \mu, L} \langle \mathcal{O}_{X} \rangle_{\beta, \mu, L}\;,
\end{equation}
as $1\ll \text{dist}(X,Y) \ll L$. The answer to this question strongly depends on the spectral properties of $\mathcal{H}$. Consider the case of weakly interacting fermionic models, Eq. (\ref{eq:H}). Suppose that $\mu$ belongs to a spectral gap of $\mathcal{H}_{0}$: 
\begin{equation}\label{eq:gapdelta}
\text{dist}( \sigma(\mathcal{H}_{0}), \mu ) \geq \delta\;,
\end{equation}
for some $\delta>0$ uniformly in $L$. For $|\lambda|$ small enough, uniformly in $\beta$ and $L$, exponential spatial decay estimates for the correlation functions can be proved thanks to the convergence of the fermionic cluster expansion, based on the Brydges-Battle-Federbush formula for fermionic truncated expectations; see {\it e.g.} \cite{BF, Bry1, GeMa, Le, Mabook, PeSa, Sa}. This result actually extends to prove exponential decay of Euclidean, or imaginary time, correlations. Let $\gamma_{t}(\mathcal{O})$ be the imaginary-time evolution of $\mathcal{O}$:
\begin{equation}\label{eq:gamma}
\gamma_{t}(\mathcal{O}) = e^{t ( \mathcal{H} - \mu \mathcal{N} )} \mathcal{O} e^{-t (\mathcal{H} - \mu \mathcal{N})}\;.
\end{equation}
In terms of the imaginary-time evolution operator, the Gibbs state satisfies the Kubo-Martin-Schwinger (KMS) identity:
\begin{equation}\label{eq:KMS}
\langle \mathcal{A} \mathcal{B} \rangle_{\beta, \mu, L} = \langle \gamma_{\beta}(\mathcal{B}) \mathcal{A} \rangle_{\beta, \mu, L}\;,
\end{equation}
which in the present finite-dimensional setting trivially follows from the cyclicity of the trace. If $\mu$ belongs to a spectral gap of $\mathcal{H}_{0}$ as in Eq. (\ref{eq:gapdelta}), the convergence of the cluster expansion for $|\lambda|$ small enough allows to prove that, for all $t\in [0,\beta)$:
\begin{equation}\label{eq:dec}
| \langle \gamma_{t}(\mathcal{O}_{X}); \mathcal{O}_{Y} \rangle_{\beta, \mu, L} | \leq Ce^{- c\text{dist}(X,Y) - c t_{\beta}}\;,
\end{equation}
with $t_{\beta} = \min(t, \beta - t)$. The constant $C$ depends on the observables, while the constant $c$ depends only on $\delta$ and on $|\lambda|$. This type of estimate can actually be used to establish the existence of a spectral gap for the many-body Hamiltonian $\mathcal{H}$, \cite{DS} for $|\lambda|$ small. More generally, in the absence of a spectral gap, the bound (\ref{eq:dec}) still holds at $\beta < \infty$, however with $\beta$-dependent constants, and with a $\beta$-dependent smallness condition on $\lambda$.

The requirement of $|\lambda|$ small is due to the fact that the proof of (\ref{eq:dec}) is based on a convergent expansion for Gibbs state of $\mathcal{H}$ around the Gibbs state of $\mathcal{H}_{0}$. If one assumes the existence of a spectral gap of $\mathcal{H}$, the analogue of (\ref{eq:dec}) for $t=0$ at zero temperature can be proved with different, non-perturbative methods, \cite{HK, NS}. Concerning decay of real-time correlation functions for many-body systems, at the moment no rigorous methods are available to prove such bounds for non-integrable many-body systems.

In the absence of a spectral gap, one does not expect exponential decay of correlations uniformly in $\beta$. Nevertheless, in some cases of physical relevance discussed later on, such as $1d$ metallic systems, and $2d$ and $3d$ semimetals, algebraic decay of ground state correlations can be proved for weak interactions. This requires the combination of fermionic cluster expansion with rigorous renormalization group methods, in order to resolve the infrared singularity that plagues naive perturbation theory around the Gibbs state of $\mathcal{H}_{0}$, and to cure it by a renormalization of physical quantities, such as the Fermi velocity or the critical exponents. 

Rigorous renormalization group methods for gapless condensed matter systems have been pioneered in \cite{BG0, FT}, see \cite{BG, Mabook, Sa} for reviews. These methods have been used to study the equilibrium state of interacting Fermi systems with extended Fermi surface at positive temperatures \cite{DR1, DR2, BGM}, exponentially small in the interaction strength, and up to zero temperature for systems with asymmetric Fermi surface \cite{FKT}. In one dimensional systems, these methods have been used to construct the interacting ground state, starting from the works \cite{BGPS, BoM}, relying on the exact integrability of the Luttinger model \cite{ML}. This property turns out to imply the vanishing of the beta function for the effective quartic coupling, which is marginal in the renormalization group sense. The integrability of the Luttinger model holds at the Hamiltonian level; such key property is broken by the presence of cut-offs, typically needed in order to set up any renormalization group analysis, and this makes the implementation of the integrability of the Luttinger model within the renormalization group analysis particularly delicate. A different strategy has been introduced in \cite{BMchiral, BMdensity,  MaQED}, where the role of integrability is replaced by the combination of Schwinger-Dyson equations and Ward identities for a regularized version of the Luttinger field theory, which allows to compute all correlation functions of the field theory and to estimate the effect of cut-offs. 

More recently, rigorous renormalization group methods have been applied to study semimetallic systems in two and three dimensions, such as graphene and Weyl semimetals, and to study transport properties at zero temperature. These extensions will be at the center of the present paper, and will be discussed in more detail below.

\section{Dynamics and linear response}\label{sec:dyn}

We are interested in the dynamical properties of the interacting lattice models defined in the previous section. To begin, notice that the Gibbs state $\rho_{\beta,\mu,L} \propto e^{-\beta (\mathcal{H} - \mu \mathcal{N})}$ is obviously invariant under the time-evolution generated by $\mathcal{H}$. In order to have a nontrivial time-evolution of the state, we have to introduce a perturbation in the Hamiltonian, and evolve the initial equilibrium state according to the new dynamics. A standard procedure is to follow an adiabatic protocol: namely, we introduce a slow time dependence in the Hamiltonian,
\begin{equation}
\mathcal{H}(\eta t) = \mathcal{H} + \varepsilon g(\eta t) \mathcal{P}\;,
\end{equation}
where: $\mathcal{P}$ is a sum of finite-range potentials, similarly to (\ref{eq:H}); $\varepsilon$ is a small parameter; $g(t)$ is a switch function, namely a smooth function such that $g(-\infty) = 0$ and $g(0) = 1$; and $\eta > 0$ allows to tune the rate of variation of the switch function. We will be interested in the situation in which $\eta$ is small, uniformly in all the other parameters in the model. A typical choice of switch function in applications is $g(t) = e^{t}$, which we shall consider from now on.

The evolution of the system is defined by the Schr\"odinger-von Neumann equation for the state:
\begin{equation}\label{eq:timeevo}
\begin{split}
i\partial_{t} \rho(t) &= [ \mathcal{H}(\eta t), \rho(t) ]\;,\qquad t\leq 0\;, \\
\rho(-\infty) &= \rho_{\beta, \mu, L}\;.
\end{split}
\end{equation}
The question addressed here is: what is the variation of the expectation values of physical observables due to the slow drive, for $t\leq 0$? To this end, it is useful to represent the variation of the expectation value of the observable $\mathcal{O}$ due to the perturbation $\mathcal{P}$ as:
\begin{equation}\label{eq:linresp}
\begin{split}
&\frac{1}{\varepsilon}\Big(\Tr \mathcal{O} \rho(0) - \Tr \mathcal{O} \rho(-\infty)\Big) \\
&= -i\int_{-\infty}^{0} dt\, e^{\eta t} \Tr \, \big[ \mathcal{O}, e^{i\mathcal{H}t} \mathcal{P} e^{-i\mathcal{H}t} \big]\rho_{\beta, \mu, L}\\&\quad + \mathcal{R}^{\beta, L}_{\mathcal{O}, \mathcal{P}}(\varepsilon, \eta) \\
&\equiv \chi^{\beta, L}_{\mathcal{O}, \mathcal{P}}(\eta) + \mathcal{R}^{\beta, L}_{\mathcal{O}, \mathcal{P}}(\varepsilon, \eta)\;,
\end{split}
\end{equation}
where the first equality follows from the Duhamel expansion in $\varepsilon$ for the non-autonomous dynamics (\ref{eq:timeevo}). The quantity $\chi^{\beta, L}_{\mathcal{O}, \mathcal{P}}(\eta)$ is the linear response of the observable $\mathcal{O}$ due to the perturbation $\mathcal{P}$, while $\mathcal{R}^{\beta, L}_{\mathcal{O}, \mathcal{P}}(\varepsilon, \eta)$ is of higher order in $\varepsilon$. In many physical applications, one is interested in computing the linear response coefficient $\chi^{\beta, L}_{\mathcal{O}, \mathcal{P}}(\eta)$ as $\beta, L\to \infty$, and possibly $\eta\to 0^{+}$. A complementary question is to prove the smallness in $\varepsilon$ of the error term $\mathcal{R}^{\beta, L}_{\mathcal{O}, \mathcal{P}}(\varepsilon, \eta)$, uniformly in $\eta$, $\beta$, $L$, which allows to rigorously justify the validity of linear response.

For gapped non-interacting systems, the validity of linear response at zero temperature, {\it i.e.} for $\beta \to \infty$, is related to the adiabatic theorem of quantum mechanics \cite{BoFo, Ka}; see \cite{Teu} for a mathematical review of adiabatic theory. The standard adiabatic theorem however does not allow to study many-body quantum systems uniformly in their size. Recently, the adiabatic theorem has been extended to the many-body setting in \cite{BdRF}, in a way that also allows to prove validity of linear reponse, see \cite{BdRFrev, HTrev} for reviews and for further references. The adaptation from quantum spin systems to fermionic models has been discussed in \cite{MT}. In the case of the quantum Hall effect, the linear response approximation can actually be proved to be exact, see \cite{KS} for non-interacting fermions and \cite{BdRFL} for the extension to many-body systems. The many-body adiabatic theorem of \cite{BdRF} applies to the setting in which the time-dependent Hamiltonian $\mathcal{H}(\eta t)$ has a non-degenerate ground state separated by the rest of the spectrum by a positive gap, for all times and uniformly in $L$. The extension to time-dependent perturbations that might close the spectral gap has been obtained in \cite{Teu2}. Finally, the extension to infinite systems, with a uniform gap or with a bulk gap, has been discussed in \cite{HenTeu1, HenTeu2}.

It is an important open problem to extend the current many-body adiabatic methods, and the validity of linear response, to positive temperature systems. The case of small systems coupled to infinite reservoirs at nonzero temperature, under suitable ergodicity assumptions on the reservoirs, has been discussed in \cite{ASF1, ASF2, JP}. Concerning the adiabatic theorem and validity of linear response for extended many-body systems, recent progress for small temperatures, vanishing as $\eta \to 0^{+}$ but uniformly in $L$, has been obtained in \cite{GLMP}, via the combination of a rigorous analytic continuation (Wick rotation) to imaginary times of the full Duhamel expansion for the non-autonomous dynamics (\ref{eq:timeevo}), and cluster expansion methods to prove the convergence of the resulting series. It is a challenging open problem to understand the interplay of the many-body adiabatic dynamics and approach to equilibrium at fixed positive temperature; see \cite{JPT1, JPT2} for further recent insights on the topic.

In the rest of this article, we will focus on rigorous results about the linear response contribution $\chi^{\beta, L}_{\mathcal{O}, \mathcal{P}}(\eta)$, for various physical systems. One of the main difficulties in studying this object is the control of the time integral, in the adiabatic limit $\eta \to 0^{+}$. In fact, it is extremely difficult to gain direct insight on the decay of real-time correlation functions for non-integrable many-body systems. One way to approach this problem is by complex deformation, by writing the transport coefficient in terms of Euclidean, or imaginary-time, correlation functions; at least for the case of weakly interacting fermionic models, Euclidean correlation functions can be estimated via fermionic cluster expansion. Let $\eta_{\beta} \in \frac{2\pi}{\beta} \mathbb{Z}$. Then, the following identity holds true, as a consequence of the KMS property (\ref{eq:KMS}) and of Cauchy theorem for analytic function:
\begin{equation}\label{eq:wick1}
\chi^{\beta, L}_{\mathcal{O}, \mathcal{P}}(\eta_{\beta}) = - \int_{0}^{\beta} ds\, e^{-i\eta_{\beta}s} \langle \gamma_{s}(\mathcal{P}) \;; \mathcal{O} \rangle_{\beta, \mu, L}\;,
\end{equation}
with $\gamma_{s}(\mathcal{P})$ the Euclidean time evolution of $\mathcal{P}$, recall (\ref{eq:gamma}). Given $\eta>0$, let $\eta_{\beta} \geq \eta$ be its best approximation in $\frac{2\pi}{\beta}\mathbb{Z}$ (obviously, $\eta_{\beta} - \eta \leq 2\pi / \beta$). Then, using only the boundedness of the fermionic operators and Lieb-Robinson bounds, see \cite{NSY} for a review, it is possible to prove that:
\begin{equation}\label{eq:wick2}
\Big| \chi^{\beta, L}_{\mathcal{O}, \mathcal{P}}(\eta_{\beta}) - \chi^{\beta, L}_{\mathcal{O}, \mathcal{P}}(\eta) \Big| \leq \frac{C}{ \eta^{d+2} \beta}\;,
\end{equation}
where the constant $C$ depends only on the support of the observable $\mathcal{O}$ (and it is independent of $L$ if $\mathcal{O}$ is local). See {\it e.g.} \cite{AMP, MPdrude} for a proof of (\ref{eq:wick1}), (\ref{eq:wick2}), and \cite{GLMP} for an extension to higher commutators. Eqs. (\ref{eq:wick1}), (\ref{eq:wick2}) allow to apply methods of Euclidean quantum field theory to the study of real-time quantum dynamics, and in particular to the evaluation of transport coefficients as $\beta \to \infty$ and $\eta \to 0^{+}$.

\section{Quantum Hall effect}\label{sec:QHE}

The quantum Hall effect is the paradigmatic example of topological transport phenomenon in condensed matter physics. It has been discovered experimentally in \cite{vK}, and the first theoretical explanation dates back to the seminal works \cite{Lau, TKNN}. The quantum Hall effect is a measurement of the charge current propagating in a $2d$ system, exposed to a transverse magnetic field and to an in-plane electric field. The measurement is performed at temperatures of the order of the Kelvin, and for a class of materials of insulating type. 

To describe the experimental observation, let us denote by $j_{i}$, $i=1,2$, the current density propagating in the two orthogonal directions on the plane. Let us denote by $E = (E_{1}, E_{2})$ the electric field. Then, neglecting higher order terms, the current in the steady state and the electric field are related by the linear response formula:
\begin{equation}
j = \sigma E\;,
\end{equation}
where $\sigma = (\sigma_{ij})_{1\leq i,j \leq 2}$ is the conductivity matrix. The entries $\sigma_{11}$, $\sigma_{22}$ are called the longitudinal conductivities, while the entries $\sigma_{12}$, $\sigma_{21}$ are called the transverse conductivities. From a phenomelogical viewpoint, the insulating behavior of the system is associated to $\sigma_{11} = \sigma_{22} = 0$: this means that transport is purely non-dissipative. Furthermore, if the system is invariant under rotations, one also has $\sigma_{12} = -\sigma_{21}$. In units such that $e^{2} = \hbar = 1$, the integer quantum Hall effect (IQHE) is the observation that the transverse, or Hall, conductivity satisfies:
\begin{equation}\label{eq:IQHE}
\sigma_{12} \in \frac{1}{2\pi} \cdot \mathbb{Z}\;,
\end{equation}
with a nine-digits experimental precision. The Hall conductivity turns out to be piecewise constant in the density of charge carries, and to display sudden jumps from one plateau to another, corresponding to a transition from different integers in (\ref{eq:IQHE}). This behavior is in glaring contrast with the expected classical behavior of $\sigma_{12}$, which in an idealized system would be linear in the density of charge carriers. See {\it e.g.} \cite{vK2} for a review of the experimental discovery of this phenomenon. Nowadays, quantum Hall systems are recognized as paradigmatic examples of topological phase of matter, see {\it e.g.} \cite{MM} for a recent review on the topic. 

The IQHE is a special case of the fractional quantum Hall effect (FQHE), which predicts the quantization of $\sigma_{12}$ in rational multiples of $1/2\pi$. In constrast to the IQHE, the theoretical understanding of the FQHE starting from microscopic models is still an outstanding problem in mathematical physics. See {\it e.g.} \cite{FB, Frev} for reviews of the effective field theoretic approach to this phenomenon. In what follows we will restrict the attention to the IQHE for lattice many-body quantum systems.

Let us consider many-body Hamiltonians of the form (\ref{eq:H}) in dimension $d=2$. We define the lattice current associated with ${\bf x}, {\bf y}$ in $\Lambda_{L}$ as:
\begin{equation}\label{eq:jxy}
j_{{\bf x}, {\bf y}} = i  a^{*}_{{\bf x}} H({\bf x}; {\bf y}) a_{{\bf y}} + \text{h.c.}\;.
\end{equation}
These operators are related to the time variation of the density operator $n_{{\bf x}} = a^{*}_{{\bf x}} a_{{\bf x}}$ via the continuity equation:
\begin{equation}\label{eq:div}
\partial_{t} \tau_{t}(n_{{\bf x}}) = \sum_{{\bf y} \in \Lambda_{L}} \tau_{t}( j_{{\bf x}, {\bf y}})\;.
\end{equation}
The right-hand side of (\ref{eq:div}) is a lattice divergence. To see this, let $Q_{X}$ be the charge operator associated with $X\subset \Lambda_{L}$, $Q_{X} = \sum_{{\bf x} \in X} n_{{\bf x}}$. Then, Eq. (\ref{eq:div}) implies that the time-variation of $Q_{X}$ is due to a flow across the boundary of $X$:
\begin{equation}
\partial_{t} \tau_{t} (Q_{X}) = \sum_{\substack{{\bf x} \in X \\ {\bf x} \notin X}} \tau_{t} (j_{{\bf x}, {\bf y}})\;.
\end{equation}
Next, we introduce the total current operator as:
\begin{equation}\label{eq:Jkdef}
\mathcal{J}_{k} = \frac{1}{2} \sum_{{\bf x}, {\bf y} \in \Lambda_{L}} (x_{k} - y_{k}) j_{{\bf x}, {\bf y}}\;,\qquad k =1,2\;,
\end{equation}
which can be viewed as $i[\mathcal{H}, \mathcal{X}_{k}]$ with $\mathcal{X}_{k}$ the second quantization of the $k$-th component of the position operator. Notice that in general the position operator is not well defined in the presence of periodic boundary conditions; nevertheless, being the Hamiltonian short-ranged, we can interpret $i[\mathcal{H}, \mathcal{X}_{k}]$ as the right-hand side of (\ref{eq:Jkdef}). The total current operator can be viewed as the value at $p=(0,0)$ of:
\begin{equation}
\hat j_{p,k} = \frac{1}{2} \sum_{{\bf x}, {\bf y} \in \Lambda_{L}} \frac{e^{ip\cdot y} - e^{ip\cdot x}}{i p\cdot (y - z)} (y - x)_{k} j_{{\bf x}, {\bf z}}\;,
\end{equation}
which satisfies the momentum-space continuity equation:
\begin{equation}\label{eq:consmom}
\partial_{t} \tau_{t}( \hat n_{p} ) = \sum_{k=1,2} i p_{k} \tau_{t} ( \hat j_{p, k} )\;.
\end{equation}
We are interested in describing the variation of the average current, after having introduced a weak electric field. The coupling with the electromagnetic field is introduced via the standard procedure used in lattice gauge theories, the Peierls' substitution:
\begin{equation}\label{eq:pei}
H({\bf x}; {\bf y}) \to H({\bf x}; {\bf y}) e^{i \int_{x\to y} d\ell \cdot A(t,\ell)}\;,
\end{equation}
where $A$ is a time-dependent vector potential, defined in $\mathbb{R}^{2}$. The integral in Eq. (\ref{eq:pei}) is taken over a straight line connecting $x$ to $y$. Let us suppose that $A(t,z)$ is constant in space, $A(t,z) \equiv A(t)$, and $A(t) = e^{\eta t} A$, so that the vector potential generates the space-homogeneous electric field $E(t) = - \eta e^{\eta t} A$. 

Let $\mathcal{H}(\eta t)$ be the time-dependent Hamiltonian, coupled to the gauge field via (\ref{eq:pei}), and let $\rho(t)$ be the solution of the Schr\"odinger-von Neumann equation (\ref{eq:timeevo}). Let $\mathcal{J}_{k}(A(t))$ the current operator in the presence of the gauge field. We are interested in computing the variation
\begin{equation}
\frac{1}{L^{2}}\Tr \mathcal{J}_{i}(A(0)) \rho(0) -  \frac{1}{L^{2}}\Tr \mathcal{J}_{i} \rho_{\beta,\mu,L}\;,
\end{equation}
at first order in the electric field. By performing one step of Duhamel expansion in the gauge field, we see that the expression for the conductivity matrix is, in the linear response approximation:
\begin{equation}\label{eq:sigmaijdef}
\begin{split}
&\sigma_{ij}^{\beta, L}(\eta) \\&= \quad\frac{i}{\eta L^{2}} \Big[ \Delta^{\beta, L}_{ij} + \int_{-\infty}^{0}ds\, e^{\eta s} \Big\langle \big[ \mathcal{J}_{i}, \tau_{s}(\mathcal{J}_{j}) \big] \Big\rangle_{\beta, \mu, L} \Big]
\end{split}
\end{equation}
where $\Delta^{\beta, L}_{i,j} = \langle [\mathcal{J}_{i}, \mathcal{X}_{j} ] \rangle$. The mathematical problem that we address is about the computation of:
\begin{equation}\label{eq:PPP}
\sigma_{ij} = \lim_{\eta \to 0^{+}} \lim_{\beta, L \to \infty} \sigma_{ij}^{\beta, L}(\eta)\;,
\end{equation}
for many-body systems with Hamiltonian (\ref{eq:H}). We will further specify that the system is in an insulating phase. To this end, it is sufficient to assume that the chemical potential $\mu$ lies in a spectral gap of the Hamiltonian, uniformly in $L$. 

Let us first briefly discuss the case of non-interacting systems. We refer the reader to \cite{AG, AW} for mathematical reviews on the topic. For $T=0$, the state of the system is completely specified by the Fermi projector $P_{\mu} = \chi(H \leq \mu)$, with $\chi(\cdot)$ the characteristic function and $\mu$ in a spectral gap of $H$. Under this assumption, the matrix elements of the Fermi projector are known to decay exponentially fast:
\begin{equation}
|P_{\mu}({\bf x}; {\bf y})| \leq Ce^{-c|x-y|}\;.
\end{equation}
The infinite volume conductivity matrix can be written as a suitable trace per unit volume via Kubo-Streda formula, see {\it e.g.} \cite{AG}:
\begin{equation}\label{eq:sigmaij}
\sigma_{ij} = \lim_{L\to \infty}\frac{i}{L^{2}} \Tr \mathbbm{1}_{L} P_{\mu} [ [ P_{\mu}, X_{i} ], [P_{\mu}, X_{j}] ]\;,
\end{equation}
where the trace is over $\ell^{2}(\mathbb{Z}^{2}; \mathbb{C}^{M})$ and $\mathbbm{1}_{L} = \chi(x \in [ -L/2, L/2 ]^{2})$. In particular, the longitudinal conductivity $\sigma_{ii}$ is zero. Instead, $\sigma_{12}$ might be non-vanishing, and it turns out to have a beautiful mathematical interpretation. At first, suppose that $H$ commutes with translations on $\mathbb{Z}^{2}$. Then, the Hamiltonian $H$ can be fibered in momentum space, as follows:
\begin{equation}
H = \int_{\mathbb{T}^{2}}^{\oplus} dk\, \hat H(k)\;,
\end{equation}
where $\hat H(k) \in \mathbb{C}^{M\times M}$ is the Bloch Hamiltonian. Similarly, the Fermi projector $P_{\mu}$ can be written as the direct integral of the fibered projectors $\hat P_{\mu}(k) = \chi(\hat H(k) \leq \mu)$. In terms of these objects, the Hall conductivity can be rewritten as an integral over the Brillouin zone:
\begin{equation}
\sigma_{12} = i \int_{\mathbb{T}^{2}} \frac{dk}{(2\pi)^{2}}\, \Tr \hat P_{\mu}(k) \Big[ \partial_{1}\hat P_{\mu}(k)\, ,\,  \partial_{2} \hat P_{\mu}(k) \Big]\;,
\end{equation}
where the trace is now over $\mathbb{C}^{M}$. The key observation \cite{AS2, TKNN} is that the argument of the integral is a curvature, and the resulting expression is equal $\frac{1}{2\pi} \cdot \mathcal{C}_{1}$, with $\mathcal{C}_{1}$ the  first Chern number of the Bloch bundle, which has $\mathbb{T}^{2}$ as base space and $\text{Ran} \hat P_{\mu}(k)$ as fibers. Thus, the Hall conductivity can only take values in $\frac{1}{2\pi} \cdot \mathbb{Z}$, and a nonzero value is associated to a nontrivial topology of the Bloch bundle. A further perspective on the triviality/nontriviality of the Bloch bundle is through the localization properties of the Wannier functions, constructed via the Bloch transform of the eigenstates of the Bloch Hamiltonian (also called Bloch functions). Informally, a vanishing Hall conductivity is equivalent to the exponential decay of the Wannier functions associated with the energy bands below the Fermi level, while a nonzero Hall conductivity is equivalent to a slow algebraic decay, implying the divergence of the squared position operator. We refer the reader to \cite{MPPT} for a precise statement and for a proof of this result.

This elegant explanation of the quantization of the Hall conductivity apparently breaks down in the absence of translation invariance. This is of course always the case in all physical applications, due to the presence of unavoidable impurities on the sample. Remarkably, the quantization of $\sigma_{12}$ survives the presence of disorder. Assuming the existence of a spectral gap for $H$, this has been proved in \cite{BES} via the methods of non-commutative geometry, and in \cite{AS2b} using functional analytic tools. Later, \cite{AG} extended the result of \cite{AS2b} to the case in which the spectral gap is replaced by a mobility gap, which is the relevant setting in the presence of Anderson localization (that is, at strong disorder), an essential ingredient to establish the existence of plateaux in the plot of $\sigma_{12}$ as a function of the density of charge carries. 

None of the mentioned rigorous results applies to interacting many-body systems, {\it e.g.} $\lambda\neq 0$ in Eq. (\ref{eq:H}). A general field-theoretic approach to the quantum Hall effect, based on the classification of effective actions of non-relativistic Fermi gases at low length scales, has been introduced in \cite{FK, F}, see \cite{FB, Frev} for reviews. This approach describes quantum Hall systems on domains with nontrivial boundaries, and it allows to understand the quantization of the Hall conductivity in terms of a cancellation mechanism between the anomaly of the Chern-Simons field theory describing the scaling limit of the bulk degrees of freedom, and the anomaly of the chiral Luttinger liquid describing the scaling limit of the edge degrees of freedom.

A first attempt in the direction of formulating the many-body conductivity matrix as a topological invariant in the spirit of \cite{TKNN} has been proposed in \cite{AS0}. There, following the original intuition of Laughlin \cite{Lau}, it is argued that the many-body Hall conductivity can be computed by probing the response of the system to a `phase twist' of the boundary conditions, physically equivalent to the introduction of a suitable magnetic flux in the system. Under an extra averaging assumption over a second magnetic flux and the assumption that the ground state of the system is gapped and non-degenerate, one can show that the many-body Hall conductivity is quantized \cite{AS0}.

The problem of removing the extra averaging assumption in \cite{AS0} remained open for a long time \cite{ASIAMP}. The proof that the extra averaging assumption can be relaxed has been given in \cite{HM}, thus giving the first proof of quantization of the Hall conductivity for a many-body quantum system up to exponentially small corrections in the size of the system, under a spectral gap assumption for the many-body ground state.

Coming back to the formulation of the conductivity matrix as linear response with respect to a constant electric field, Eq. (\ref{eq:sigmaijdef}), a different route to quantization is to show that Eq. (\ref{eq:PPP}) is actually constant in the strength of the many-body interaction, provided the spectral gap of $\mathcal{H}$ does not close along the path that connects the Hamiltonian to its noninteracting counterpart ($\lambda = 0$). The persistence of the spectral gap is a consequence of the convergence of the fermionic cluster expansion, which can be proved for $|\lambda|$ small enough, and which allows to prove exponential decay of correlations (\ref{eq:dec}). After proving that the conductivity matrix stays constant along the path that deforms the model into a non-interacting one, quantization of $\sigma_{12}$ follows from the known results for non-interacting systems. This approach to quantization has been introduced in \cite{GMPhall}, for translation-invariant systems, and it has been extended in \cite{GJMP, GMPhald} to study the topological phase transition of the Haldane-Hubbard model, including the construction of an interacting critical line across which $\sigma_{12}$ has a jump discontinuity. The starting point of the approach is the rigorous Wick rotation, that allows to represent the conductivity matrix in terms of Euclidean correlation functions, recall Eqs. (\ref{eq:wick1}), (\ref{eq:wick2}). Then, the constancy of the conductivity matrix along the path of Hamiltonians that deforms $\mathcal{H}$ into $\mathcal{H}_{0}$ is proved using lattice Ward identities, following from the conservation of lattice current (\ref{eq:consmom}), in a way inspired by the non-renormalization of the so-called topological mass in QED3, \cite{CH}. 

Another approach to quantization of $\sigma_{12}$ has been proposed in \cite{BBdRF}, where the quantization of the Hall conductivity for gapped many-body systems is proved via a many-body index theorem for the ground state projector. Assuming a suitable degeneracy condition for the many-body ground state, the result of \cite{BBdRF} also shows fractional quantization of the response coefficient. Proving the required degeneracy assumption on the many-body projectors starting from an interacting microscopic model is a difficult open problem in mathematical physics.

All the above results have been obtained for many-body systems under a spectral gap assumption. A limitation of this setting is that it does not allow to prove the emergence of plateaux in the plot of the Hall conductivity as a function of the density of charge carriers. As for the non-interacting case, in order to understand the emergence of plateaux, one should include strong disorder in the microscopic model. It is an important problem to understand the interplay of interactions and disorder effects in many-body quantum Hall systems.

\section{Transport in semimetals}\label{sec:semi}

So far, we have focused on two-dimensional gapped many-body systems, for which transport is non-dissipative, and where the transverse conductivity exhibits interesting quantization properties. In the absence of a spectral gap, it turns out that the transport properties of the system strongly depend on the nature of the low-energy excitations at the Fermi level. In this section we shall consider translation-invariant models, with Bloch Hamiltonian $\hat H(k) \in \mathbb{C}^{M\times M}$, $k\in \mathbb{T}^{d}$. Given a value of the chemical potential $\mu$, the Fermi surface is defined as:
\begin{equation}
\Big\{ k\in \mathbb{T}^{d} \,\, \big|\,\, \mu \in \sigma(\hat H(k)) \Big\}\;.
\end{equation}
An insulator corresponds to the situation in which the Fermi surface is empty. Instead, a metallic system corresponds to the situation in which the Fermi surface has co-dimension $1$. The standard example is the one of the lattice Laplacian $-\Delta_{\mathbb{Z}^{d}}$ and $\mu$ chosen within the spectrum of the Laplacian. For this class of systems, the zero-temperature conductivity matrix is typically infinite.

Insulators and metals do not exhaust all possible physical situations. Semimetals are a class of physical systems for which the Fermi surface is nonempty but nevertheless the conductivity matrix is finite. Here we shall review the cases of graphene and of Weyl semimetals.
\subsection{Graphene}
Graphene is a paradigmatic example of semimetal in two dimensions, consisting of a one-atom thick layer of graphite \cite{GN}. Its Fermi surface in the absence of doping is given by two points, $\{ k_{F}^{+}, k_{F}^{-} \}$. Around the two Fermi points, the dispersion relation of graphene takes a peculiar shape, mimicking the dispersion relation of $2+1$ dimensional massless Dirac fermions, which is responsible for its remarkable transport properties \cite{Kat}.

The simplest tight-binding model for graphene is the Laplacian on the honeycomb lattice, \cite{W}. The honeycomb lattice can be viewed as the superposition of two triangular sublattices, generated by the basis vectors $a_{1} = (1/2)( 3, \sqrt{3} )$, $a_{2} = (1/2)( 3, -\sqrt{3} )$:
\begin{equation}
\begin{split}
\Lambda_{L}^{A} &= \Big\{ x_{1}a_{1} + x_{2} a_{2}\, \Big|\, 0\leq x_{i} \leq L-1 \Big\}\;,\\
\Lambda_{L}^{B} &= \Lambda_{L}^{A} + (1,0)\;.
\end{split}
\end{equation}
with edges connecting sites of $\Lambda^{A}_{L}$ with sites of $\Lambda^{B}_{L}$ at distance one. Equivalently, the vertex set $\Lambda_{L}^{A} \cup \Lambda^{B}_{L}$ of the honeycomb lattice can be thought of as the union over $x\in \Lambda_{L}^{A}$ of the pairs of sites $(x, x+(1,0))$; that is, as a decorated triangular lattice with two internal degrees of freedom. The advantage of this representation is that, in contrast to the honeycomb lattice, the triangular lattice is a Bravais lattice. Thus, given $x = (x_{1}, x_{2}) \in \Lambda_{L}^{A}$, the single-particle wave function $\psi(x)$ is a two-component function, collecting values in the two shifted sublattices labelled by $x$, $\psi(x) = (\psi_{A}(x), \psi_{B}(x + (1,0)))$. 

The Bloch Hamiltonian $\hat H(k)$ of the Laplacian on the honeycomb lattice is:
\begin{equation}\label{eq:HB}
\hat H(k) = -t \begin{pmatrix} 0 & \Omega(k) \\ \Omega(k)^{*} & 0 \end{pmatrix}\;,
\end{equation}
where $t > 0$ is the hopping parameter and the complex-valued function $\Omega(k)$ is given by:
\begin{equation}
\Omega(k) = 1 + e^{-ik_{1}} + e^{-ik_{2}}\qquad k\in \mathbb{T}^{2}\;.
\end{equation}
The energy bands of the model are $\pm t |\Omega(k)|$, and they touch at the two inequivalent Fermi points on $\mathbb{T}^{2}$:
\begin{equation}
k_{F}^{+} = \Big( \frac{2\pi}{3}, \frac{4\pi}{3} \Big) \;,\qquad k_{F}^{-} = \Big(\frac{4\pi}{3}, \frac{2\pi}{3}\Big) \;.
\end{equation}
Taking into account the two spin degrees of freedom, the half-filling condition corresponds to the choice $\mu = 0$, for which the Fermi surface is given by $k_{F}^{+}$ and by $k_{F}^{-}$. This choice of chemical potential is of course non-generic, but it is the physically relevant one to describe neutral graphene. In proximity of the Fermi surface, the Bloch Hamiltonian can be approximated by the expression
\begin{equation}
\begin{split}
\hat H(k' + k_{F}^{\omega}) &= -\frac{3t}{2}\begin{pmatrix} 0 & ik'_{1} - \omega k'_{2} \\ -ik'_{1} - \omega k'_{2} & 0  \end{pmatrix}\\&\quad + O(|k'|^{2})\;,
\end{split}
\end{equation}
which mimics, at low energy, the Hamiltonian of $2+1$ dimensional massless Dirac fermions. We shall denote by $\langle \cdot \rangle_{\beta, L} \equiv \langle \cdot \rangle_{\beta, 0, L}$ the corresponding Gibbs state. 

In the absence of interactions, the Gibbs state of the system, and more generally the Euclidean correlation functions of the system, are completely specified by the two-point function,
\begin{equation}\label{eq:2pt}
\big\langle {\bf T}\, \gamma_{t}( a^{+}_{{\bf x}} ) \gamma_{s}(a^{-}_{{\bf y}}) \big\rangle^{0}_{\beta, L}\qquad t,s\in [0,\beta)\;,
\end{equation}
with ${\bf T}$ the fermionic imaginary-time ordering and where ${\bf x}, {\bf y} \in \Lambda_{L}^{A} \times \{A, B\}$ label points on the honeycomb lattice, understood as points on a decorated triangular lattice. In the absence of interactions the Gibbs state is quasi-free, and hence higher order, time-ordered Euclidean correlations can be computed from (\ref{eq:2pt}) by application of the fermionic Wick's rule. Let us focus on the ground state properties of the system, in the infinite volume limit. We denote by $\langle \cdot \rangle^{0}_{\infty} = \lim_{\beta \to \infty} \lim_{L\to \infty} \langle \cdot \rangle^{0}_{\beta, L}$ the (thermal) ground state average for the non-interacting system. Due to the absence of a spectral gap at the Fermi level, it turns out that the ground state correlation functions have a non-integrable space-time decay:
\begin{equation}\label{eq:2pt2}
\Big|\big\langle {\bf T}\, \gamma_{t}( a^{+}_{{\bf x}} ) \gamma_{s}(a^{-}_{{\bf y}}) \big\rangle^{0}_{\infty}\Big| \sim \frac{1}{\| x - y \|^{2} + | t - s|^{2} }\;.
\end{equation} 
This asymptotic behavior can be easily checked from an explicit computation, starting from the Bloch Hamiltonian (\ref{eq:HB}). 

Let us now focus on many-body models for graphene. The same qualitative algebraic decay of correlations can be proved for weakly interacting graphene \cite{GM}, with Hamiltonian $\mathcal{H} = \mathcal{H}_{0} + \lambda \mathcal{V}$, where $\mathcal{H}_{0}$ is the second quantization of the lattice Laplacian, $\mathcal{V}$ is a short-ranged many-body Hamiltonian, and $|\lambda|$ is small. The analysis of \cite{GM} is based on rigorous renormalization group methods, and on the convergence of the fermionic cluster expansion. The universality of the critical exponents of graphene in the presence of weak short range interactions ultimately relies on the fact that short-range interactions are irrelevant in the renormalization group sense for this class of systems. Nevertheless, many-body interactions do affect the large scale behavior of the system, in the sense of a non-trivial wave function renormalization and a non-trivial renormalization of the Fermi velocity. 

The slow decay of correlations has remarkable consequences on the transport properties of the system. Let $\langle \cdot \rangle_{\beta, L}$ be the Gibbs state of weakly interacting graphene. We shall focus on the conductivity matrix of the system, following \cite{GMPgra}. Let $\sigma^{\beta, L}_{ij}(\eta_{\beta})$ be the conductivity matrix of the system (\ref{eq:sigmaijdef}) after Wick's rotation:
\begin{equation}\label{eq:condgra}
\begin{split}
&\sigma^{\beta, L}_{ij}(\eta_{\beta}) =\\&\frac{1}{\eta_{\beta} L^{2} A}\Big[ i\Delta^{\beta, L}_{ij} + \int_{-\beta/2}^{\beta/2} ds\, e^{-i\eta_{\beta} s} \langle {\bf T} \mathcal{J}_{i}\;; \gamma_{s}(\mathcal{J}_{j}) \rangle_{\beta, L} \Big]\;,
\end{split}
\end{equation}
where $A = 3\sqrt{3} / 2$ is the area of the lattice fundamental cell. Let $\sigma_{ij}(\eta) = \lim_{\beta\to \infty} \lim_{L\to \infty} \sigma^{\beta, L}_{ij}(\eta_{\beta})$ be the ground-state conductivity matrix. Lattice Ward identities, following from the conservation of the lattice current (\ref{eq:consmom}), can be used to express $\sigma_{ij}(\eta)$ in terms of the Euclidean current-current correlation function \cite{GMPgra}:
\begin{equation}\label{eq:wickcond}
\begin{split}
&\sigma_{ij}(\eta) =\\
&\lim_{\beta, L \to \infty}\frac{1}{\eta_{\beta} L^{2} A}\Big[ \int_{-\frac{\beta}{2}}^{\frac{\beta}{2}} ds\, \Big(e^{-i\eta_{\beta} s} - 1\Big) \langle {\bf T} \mathcal{J}_{i}\;; \gamma_{s}(\mathcal{J}_{j}) \rangle_{\beta,L}\;.
\end{split}
\end{equation}
The expression (\ref{eq:wickcond}) is the starting point of a renormalization group analysis of the conductivity matrix of graphene. A key role in the analysis is played by the function:
\begin{equation}\label{eq:Kij}
K_{ij}(\eta) = \lim_{\beta, L \to \infty}\frac{1}{L^{2} A} \int_{-\frac{\beta}{2}}^{\frac{\beta}{2}} ds\, e^{-i\eta_{\beta} s} \langle {\bf T} \mathcal{J}_{i}\;; \gamma_{s}(\mathcal{J}_{j}) \rangle_{\beta,L}
\end{equation}
in terms of which we can rewrite (\ref{eq:wickcond}) as:
\begin{equation}\label{eq:lim}
\sigma_{ij} = \lim_{\eta \to 0^{+}} \frac{1}{\eta} \Big( K_{ij}(\eta) - K_{ij}(0) \Big)\;.
\end{equation}
The off-diagonal part of the conductivity matrix turns out to be zero, by time-reversal symmetry \cite{GMPgra}. Instead, the diagonal part of the conductivity matrix, called the longitudinal conductivity, is expected to be non-zero: this is related to the presence of a non-zero value of the minimal conductivity in graphene \cite{Na}. Remarkably, as measured in \cite{Na}, the minimal longitudinal conductivity of graphene appears to be universal:
\begin{equation}\label{eq:nair}
\sigma_{ii} = \frac{e^{2}}{h} \frac{\pi}{4}\;,
\end{equation}
or equivalently $\sigma_{ii} = 1/8$ in units such that $e^{2} = \hbar = c = 1$. Thus, up to very good precision, the longitudinal conductivity appears to depend only on fundamental constants. 

Concerning theoretical predictions, for non-interacting systems the value (\ref{eq:nair}) can be obtained via an explicit computation \cite{SPG}, starting from the model (\ref{eq:HB}). Remarkably, the expression found in \cite{SPG} agrees with the analogous quantity computed for $2+1$ dimensional massless Dirac fermions \cite{Seme}.

The main challenge is thus to understand the universality of $\sigma_{ii}$ for many-body models describing interacting graphene. This has been achieved in \cite{GMPgra}; let us sketch the strategy of the proof. Coming back to Eq. (\ref{eq:lim}), this expression shows that if the function that $K_{ij}(\eta)$ is differentiable at $\eta = 0$, then $\sigma_{ij}$ would simply be the derivative at $\eta = 0$ of $K_{ij}(0)$. At the same time, it is not difficult to see that $K_{ii}(\eta)$ is even in $\eta$ \cite{GMPgra}. Thus, this symmetry combined with differentiability would imply a trivial longitudinal conductivity, giving a result in contradiction with the experimental observation \cite{Na}.

The key point is that, due to the slow decay of correlations, already visible in the absence of interactions (\ref{eq:2pt2}), the function $K_{ij}(\eta)$ is continuous but not differentiable at zero. A careful analysis of the renormalized expansion for the current-current correlations of weakly interacting graphene allows to isolate the singular contribution to $K_{ii}(\eta)$ from its regular part:
\begin{equation}
K_{ii}(\eta) = K^{\text{Dirac}}_{ii}(\eta) + K_{ii}^{\text{R}}(\eta)\;,
\end{equation}
where $K^{\text{Dirac}}_{ii}$ is the quantum field theory analogue of (\ref{eq:Kij}), associated with the integration over a non-interacting $2+1$ dimensional massless Dirac field, with renormalized Fermi velocity and wave function, and with a fixed ultraviolet cutoff. Instead, the remainder term $K^{\text{R}}(\eta)$ is much less explicit, and it collects irrelevant contributions generated in the renormalization group analysis, such as those taking into account the nonlinearity of the energy band away from the Fermi points. Crucially, the term $K_{ii}^{\text{R}}(\eta)$ turns out to be differentiable at $\eta = 0$. By inspection, $K^{\text{Dirac}}_{ii}(\eta)$ is even in $\eta$, which ultimately implies that $K_{ii}^{\text{R}}(\eta)$ is also even, and hence, by its improved regularity, it does not contribute to $\sigma_{ii}$. We thus find:
\begin{equation}\label{eq:sigmaD}
\sigma_{ii} = \lim_{\eta \to 0^{+}} \frac{1}{\eta} \Big( K^{\text{Dirac}}_{ij}(\eta) - K^{\text{Dirac}}_{ij}(0) \Big)\;.
\end{equation}
The right-hand side of (\ref{eq:sigmaD}) is now amenable to an explicit computation. The remaining interaction dependence, due to the finite renormalizations of the Dirac propagator, turns out to cancel out exactly, which allows to prove the validity of (\ref{eq:nair}) for weakly interacting graphene \cite{GMPgra}.

\subsection{Weyl semimetals}

Weyl semimetals are three-dimensional electron systems, whose low energy properties are well described by $3+1$ dimensional massless Weyl fermions, in the same way in which massless Dirac fermions emerge in graphene. 

The simplest setting one can use to model Weyl semimetals is the one of non-interacting, three dimensional lattice fermions on a bipartite lattice, with Bloch Hamiltonian $\hat H(k)$ on $\mathbb{T}^{3}$, such that
\begin{equation}\label{eq:weyl}
\hat H(k' + k_{F}^{\omega}) = v_{1} \sigma_{1} k'_{1} + v_{2} \sigma_{2} k'_{2} + \omega v_{3} \sigma_{3} k'_{3} + O(|k'|^{2})\;,
\end{equation}
where $k_{F}^{\omega} = (0, 0, \omega k_{F,3})$ with $\omega = \pm$. The points $k_{F}^{\pm}$ are called Weyl points. In Eq. (\ref{eq:weyl}), $\sigma_{i}$, $i=1,2,3$, are the standard Pauli matrices, and the parameters $v_{i}$ play the role of emergent velocities for the low-energy excitations. A concrete example of microscopic lattice model with Bloch Hamiltonian satisfying Eq. (\ref{eq:weyl}) has been introduced in \cite{DLC}. Eq. (\ref{eq:weyl}) simulates a relativistic Hamiltonian at energies in proximity of $\mu = 0$. The two Weyl points introduce two effective chiralities for the low energy excitations around $\mu = 0$, while the two sublattices play the role of spin degrees of freedom for the emergent relativistic field. 

One of the most peculiar phenomena of QED in $3+1$ dimensions is the presence of the chiral anomaly. Consider the Lagrangian of QED for a massless field $\psi$ coupled with a gauge field $A_{\mu}$,
\begin{equation}
\mathcal{L}(\psi, A) = \int dx\, \overline{\psi}_{x} \gamma^{\mu} (i \partial_{\mu} - A_{\mu, x}) \psi_{x}\;.
\end{equation}
At the classical level, the invariance of the Lagrangian under $U(1)$ gauge transformations and under chiral gauge transformations implies the following conservation laws, by Noether's theorem:
\begin{equation}\label{eq:cons}
\partial_{\mu} j_{\mu} = 0\;,\qquad \partial_{\mu} j^{5}_{\mu} = 0\;,
\end{equation}
with $j_{\mu} = \overline{\psi} \gamma_{\mu} \psi$ the current and $j^{5}_{\mu} = \overline{\psi} \gamma_{\mu} \gamma_{5} \psi$ the chiral current. It is well-known that these conservation laws might be broken in the quantization of the gauge theory, because of the unavoidable presence of cutoffs; these are needed to make sense of the theory, and might break the gauge symmetries of the model. The typical question is whether the conservation laws are restored after removing the cutoffs. As discovered in \cite{ADL, BJ}, this is not the case for the conservation of the chiral current. One has, in momentum space, and in units such that $e^{2} = \hbar = c = 1$:
\begin{equation}\label{eq:ABJ}
\begin{split}
&( p_{1,\mu} + p_{2,\mu} ) \langle j^{5}_{\mu, p_{1} + p_{2}} \;; j_{\nu, -p_{1}}\;; j_{\sigma, -p_{2}} \rangle^{QED} \\&\qquad = -\frac{1}{2\pi^{2}} \varepsilon_{\alpha\beta\nu\sigma} p_{1,\alpha} p_{2,\beta}
\end{split}
\end{equation}
where $\varepsilon$ is the totally antisymmetic tensor, and $\langle \cdot \rangle^{QED}$ denotes the average taken over the QED's path integral, expressed in terms of a (formal) sum over connected Feynman diagrams, after renormalization. The fact that the right-hand side of (\ref{eq:ABJ}) is nonzero implies that the classical conservation law of the chiral current in (\ref{eq:cons}) does not survive quantization: it is an example of anomaly in quantum field theory. The right-hand side of (\ref{eq:ABJ}) takes an extremely simple form, and it is manifestly not affected by radiative corrections. This cancellation has been first discussed, order by order in perturbation theory, in \cite{AB}, showing that the only contribution to the right-hand side of (\ref{eq:ABJ}) arises from the so-called `triangle graph'. The physical consequence of the chiral anomaly (\ref{eq:ABJ}) is the decay of the neutral pion into photons.

In the condensed matter setting, the counterpart of the chiral anomaly (\ref{eq:ABJ}) has been first discussed in \cite{NN}. There, it was predicted that, for three dimensional lattice models displaying the relativistic structure (\ref{eq:weyl}), a condensed matter counterpart of the chiral anomaly in QED should take place, after coupling the system to an external time-dependent gauge field. Calling $\dot{N}_{+}$ and $\dot{N}_{-}$ the time variations of the charge around the Fermi points with $\omega = +$, resp. $\omega = -$, the prediction of \cite{NN} is that:
\begin{equation}\label{eq:NN}
\dot{N}_{+} - \dot{N}_{-} = \frac{1}{2\pi^{2}} E\cdot B\;.
\end{equation}
where $E$ and $B$ are the electric and magnetic field associated to the external gauge field.  Eq. (\ref{eq:NN}) has to be understood as a steady charge flow in momentum space from one Weyl node to the other, which mimics the creation and annihilation of particles with opposite chiralities in QED. In the condensed matter setting, this phenonenon is made possible by the fact that the two Weyl nodes are connected via the energy bands. This transport phenomenon has been experimentally observed in the last years, we refer the reader to \cite{AMV} for a review on Weyl semimetals and for their transport properties.

We shall now describe the mathematical analysis of the phenomenon, following \cite{GMPweyl}. We shall consider weakly interacting Weyl semimetals, with Hamiltonian $\mathcal{H} = \mathcal{H}_{0} + \lambda \mathcal{V}$, where $\mathcal{H}_{0}$ is the second quantization of a single-particle Hamiltonian with the properties (\ref{eq:weyl}), and $\mathcal{V}$ is a short-ranged many-body interaction. We denote by $\langle \cdot \rangle_{\beta, L}$ the Gibbs state of such Hamiltonian, in the presence of a suitably chosen staggered chemical potential that ensures the existence of two Weyl nodes for the interacting system as well. 

The first non-trivial issue is to identify an operator which allows to capture the chiral charge transfer. This is nontrivial, because the charge of the emergent Weyl fermions cannot be defined exactly using local operators. Nevertheless, one can identify operators whose correlation functions behave in the desired way, at large scales. For instance, we consider:
\begin{equation}
n^{5}_{{\bf x}} = \frac{i Z^{5}}{2} \big( a^{*}_{{\bf x}} a_{{\bf x} + e_{3}} - a^{*}_{{\bf x} + e_{3} }a_{{\bf x}}\big)\;,
\end{equation}
with $e_{3} = (0, 0, 1)$. The parameter $Z^{5}$ is a suitable normalization, that has to be fixed consistently with the renormalization of the usual charge. Observe that, setting $N^{5} = \sum_{{\bf x}} n^{5}_{{\bf x}}$,
\begin{equation}\label{eq:N5}
N^{5} = Z^{5} \int \frac{d k}{(2\pi)^{2}} \sin k_{3} \hat n_{k}\;,
\end{equation}
with $\hat n_{k}$ the density operator in momentum space. In the integral (\ref{eq:N5}), thanks to the presence of $\sin k_{3}$,  the contribution due to the momenta close to $k_{F}^{\pm}$ mimics the difference of charges with opposite chiralities. The parameter $Z^{5}$ is fixed imposing the equality of the chiral vertex function with the charge vertex function, at low energy:
\begin{equation}
\langle \hat n^{5}_{p}; \hat a_{k+p}; \hat a^{*}_{k} \rangle_{\infty} = \omega \langle \hat n_{p}\;;\hat a_{k+p}\;; \hat a^{*}_{k} \rangle_{\infty} (1 + o(1))\;,
\end{equation}
where $o(1)$ is a quantity that vanishes for $k \to k_{F}^{\omega}$ and $p\to 0$.

We couple the Hamiltonian $\mathcal{H}$ to an external gauge potential $e^{\eta t}A_{\mu,x}$, via the Peierls' substitution (\ref{eq:pei}), with $\eta > 0$ and $t\leq 0$, and we denote by $\mathcal{H}(\eta t)$ the corresponding time-dependent Hamiltonian.  We are interested in the quadratic variation in the gauge field for the evolution of the expectation value of $N^{5}$. Notice that, being the observable $N^{5}$ nonlocal, it also couples to the gauge field via the introduction of a Wilson line, similarly to (\ref{eq:pei}). 

Let $\rho(t)$ be the solution of the Schr\"odinger equation (\ref{eq:timeevo}). We consider the quadratic response,
\begin{equation}\label{eq:respweyl}
\begin{split}
&\Big[ \partial_{t} \Tr N^{5}(A(t)) \rho(t) \Big]^{(2)} \\&= \int \frac{dp}{(2\pi)^{3}} \hat A_{\mu, p} \hat A_{\nu, -p} \eta \Gamma^{5; \beta, L}_{\mu, \nu}((\eta, p), (\eta, -p))\;,
\end{split}
\end{equation}
and the goal is to compute the response function $\Gamma^{5}_{\mu, \nu}$. The result of \cite{GMPweyl} is that, for $|\lambda|$ small enough, setting $\Gamma^{5} = \lim_{\beta \to \infty} \lim_{L \to \infty} \Gamma^{5; \beta, L}$:
\begin{equation}\label{eq:weylanomaly}
\begin{split}
&2\eta \Gamma_{\mu,\nu}^{5}((\eta, p), (\eta, -p)) \\ &= -\frac{1}{2\pi^{2}} p_{1,\alpha} p_{2,\beta} \varepsilon_{\alpha\beta\mu\nu} + o(|p_{i}|^{2})
\end{split}
\end{equation}
where $p_{1} = (\eta, p)$ and $p_{2} = (\eta, -p)$ are Euclidean momenta with four components. Eq. (\ref{eq:weylanomaly}), combined with (\ref{eq:respweyl}), agrees with the prediction (\ref{eq:NN}), with the same prefactor. As for the universality of graphene's conductivity, the method of proof relies on rigorous renormalization group methods, allowing us to isolate the scaling limit contribution to $\Gamma_{\mu,\nu}^{5}$, which can be computed explicitly in terms of the renormalized triangle graph, from a remainder term, which is not explicit but enjoys better regularity properties. Universality follows from the combination of lattice Ward identities, due to the conservation of the lattice current, with emergent Ward identities, valid for the correlation functions of QED. Eq. (\ref{eq:weylanomaly}) can be viewed as a rigorous version of the non-renormalization mechanism of Adler and Bardeen \cite{AB} for QED, for a short-ranged lattice model. Remarkably, the non-renormalization of the lattice counterpart of the chiral anomaly survives the breaking of rotation symmetry due to the lattice. The proof relies on lattice conservation laws, and holds at a nonperturbative level.

\section{Transport in $1d$ and quasi-$1d$ metallic systems}\label{sec:1d}

In the previous section we discussed condensed matter systems whose Fermi surface is formed by isolated points, around which the energy dispersion relation displays a typical `relativistic' shape. The Gibbs state in the presence of weak many-body interactions can be constructed rigorously, via renormalization group methods. For these models, many-body interactions of order four and higher turn out to be irrelevant in the renormalization group sense, and the large scale behavior for the system is effectively described in terms of a non-interacting relativistic field, whose covariance is renormalized by the integration of the degrees of freedom on smaller scales.

Here we discuss the transport properties of one-dimensional lattice models, where the low energy excitations are again described by emergent relativistic fields, but where the quartic many-body interaction turns out to be marginal in the renormalization group sense. This means that the large scale behavior of the system is described by an interacting quantum field theory. The analysis of the emergent quantum field theory is facilitated by its special integrability features, which are typical of the one-dimensional world. In the following, we will focus on the response functions of one-dimensional quantum systems, and of the edge currents of two-dimensional quantum Hall systems.

\subsection{Interacting $1d$ systems}

Let us consider interacting spinless fermions on $\Lambda_{L} = [-L/2, L/2] \cap \mathbb{Z}$ with periodic boundary conditions. The many-body Hamiltonian is $\mathcal{H} = \mathcal{H}_{0} + \lambda \mathcal{V}$ with:
\begin{equation}\label{eq:H1d}
\begin{split}
\mathcal{H}_{0} &= -t\sum_{x\in \Lambda_{L}} (a^{*}_{x} a_{x+1} + a^{*}_{x+1} a_{x}) \\
\mathcal{V} &= \sum_{x, y\in \Lambda_{L}} a^{*}_{x}a_{x} v(x-y) a^{*}_{y} a_{y}\;,
\end{split}
\end{equation}
with $v(x-y)$ finite-ranged and $t>0$. The Gibbs state of the model is denoted by $\langle \cdot \rangle_{\beta, \mu, L}$, where $\mu$ is chosen within the spectrum of the infinite volume lattice Laplacian, $\mu \in (-2t, 2t)$. For the special choice of nearest-neighbour interactions, the Hamiltonian $\mathcal{H}$ is also related to the Hamiltonian of the XXZ quantum spin chain, via the Jordan-Wigner transformation, and it is therefore integrable via Bethe ansatz. For generic choices of the interaction potential, however, the model is not integrable. Unless otherwise stated, in the following we will not assume any special integrability property.

We shall focus on the transport properties of the density and the current operator, defined as:
\begin{equation}
n_{x} = a^{*}_{x} a_{x}\;,\qquad j_{x} = -i t (a^{*}_{x} a_{x+1} - a^{*}_{x+1} a_{x})\;,
\end{equation}
related by the lattice continuity equation
\begin{equation}
\partial_{t} \tau_{t}(n_{x}) + \text{d}_{x} \tau_{t}(j_{x}) = 0\;,
\end{equation}
with $\text{d}_{x}$ the discrete derivative, $\text{d}_{x} f(x) = f(x) - f(x-1)$. 

We will be interested in the variation of the density and of the current operators, after introducing a slowly varying external potential $e^{\eta t} \mu(x)$, or a slowly varying electromagnetic field generated by an external vector potential $e^{\eta t} A_{x}$, compatible with the periodicity of $\Lambda_{L}$. That is, the quantities of interest will be:
\begin{equation}\label{eq:nj}
\Tr n_{x} \rho(t)\;,\qquad \Tr j_{x} \rho(t)\;,
\end{equation}
where $\rho(t)$ is the solution of the Schr\"odinger-von Neumann equation, with time-dependent Hamiltonian:
\begin{equation}
\begin{split}
\mathcal{H}(\eta t) &= \mathcal{H} + \varepsilon e^{\eta t} \sum_{x\in \Lambda_{L}} \mu(x) n_{x}\quad \text{(for the density)}\\
\mathcal{H}(\eta t) &= \mathcal{H}_{0}(A_{t}) + \lambda \mathcal{V}\quad \text{(for the current)}\;,
\end{split}
\end{equation}
where $\mathcal{H}_{0}(A_{t})$ is the free Hamiltonian coupled with an external vector potential $e^{\eta t}A_{x}$, via the Peierls substitution (\ref{eq:pei}). We define the susceptibility $\kappa_{\beta, L}(\eta, p)$ and the Drude weight $D_{\beta, L}(\eta, p)$ as the linear response of the average density and of the average current:
\begin{equation}\label{eq:kappaD}
\begin{split}
&\kappa_{\beta, L}(\eta, p) \\& \quad = \frac{i}{L} \int_{-\infty}^{0} dt\, e^{\eta t} \langle [ \hat \rho_{-p}, \tau_{t}(\hat \rho_{p}) ] \rangle_{\beta, \mu, L} \\
&D_{\beta, L}(\eta, p) \\ &\quad = \frac{-i}{L} \Big( \int_{-\infty}^{0}dt\, e^{\eta t} \langle [ \hat j_{-p}, \tau_{t}(\hat j_{p}) ] \rangle_{\beta, \mu, L} + \Delta_{\beta, L}\Big)\;,
\end{split}
\end{equation}
with $\Delta^{\beta, L} = \langle [ \mathcal{J}, \mathcal{X} ] \rangle_{\beta, \mu, L}$ (compare with Eq. (\ref{eq:sigmaijdef})). In terms of these quantities, one can write the linear response of the current and of the density in (\ref{eq:nj}) as:
\begin{equation}
\begin{split}
\Tr\, n_{0} \rho(0) &= \frac{1}{L} \sum_{p}\hat \mu(p) \kappa_{\beta, L}(\eta, p) + O(\mu^{2}) \\
\Tr\, j_{0} \rho(0) &= \frac{1}{L} \sum_{p} \hat A_{p} D_{\beta, L}(\eta, p) + O(A^{2})\;.
\end{split}
\end{equation}
The Drude weight is related to the conductivity of the model, defined as $\sigma_{\beta, L}(\eta, p) = (1/\eta) D_{\beta, L}(\eta, p)$. Thus, a finite Drude weight as $\eta \to 0^{+}$ implies a divergent conductivity, {\it i.e.} metallic behavior. Several alternative definitions of the Drude weight exists, see \cite{MPdrude} for a review; we shall refer to the expression in (\ref{eq:kappaD}) as the canonical Drude weight.

As mentioned above, for nearest-neighbour interactions the Hamiltonian (\ref{eq:H1d}) can be mapped to the XXZ model, which is exactly solvable via Bethe ansatz \cite{YY}. As a consequence, in this special case the Drude weight and the susceptibility can be explicitly computed at zero temperature. Much less is known at positive temperature. There, the Drude weight of the XXZ chain can be proved to be bounded below by a nonzero quantity \cite{Zotos} by the Mazur bound \cite{Maz}. More generally, it has been proposed that the Drude weight of many-body one dimensional systems is nonzero or zero depending on whether the model is integrable or not; we refer the reader to \cite{Pdrude} for a review of this topic.

The question addressed here, which is amenable to a rigorous analysis, is about the computation of the Drude weight and the susceptibility at zero temperature, for non-integrable many-body quantum systems. Heuristic insight on the value of such transport coefficients can be gained from the comparison of the original lattice model (\ref{eq:H1d}) with its scaling limit, the Luttinger model \cite{Luttinger}. The Hamiltonian of this one-dimensional model is:
\begin{equation}\label{eq:HL}
\begin{split}
\mathcal{H}_{\text{L}} &= \sum_{\omega = \pm} \int dx\, \overline{\psi}_{\omega, x} i c_{\omega} \partial_{x} \psi_{\omega, x} \\&\quad + \sum_{\omega} \int dxdy\, \overline{\psi}_{+, x} \psi_{+, x} w(x-y) \overline{\psi}_{-, y} \psi_{-, y}
\end{split} 
\end{equation}
where $\overline{\psi}$, $\psi$ are conjugate fermionic fields, with chirality $\omega = \pm$, and velocities $c_{\omega} = \omega c$, and where $w(\cdot)$ is a smooth and short-ranged potential. This model arises from the linearization of the dispersion relation $\varepsilon(k) = -2t\cos(k)$ around the two Fermi points $k_{F}^{\pm}$, which are solutions of the equation $\mu = -2t\cos(k_{F}^{\omega})$. It is well-known that the Luttinger model can be solved by bosonization, \cite{ML}: it can be mapped into a non-interacting bosonic quantum field theory, and the main effect of the interaction is to remove the discontinuity in the occupation number of the fermionic modes in the ground state. More generally, the many-body interaction gives rise to the appearence of interaction-dependent anomalous exponents in the scaling of the correlation function. Notice that with respect to the original Luttinger model, here we are considering a many-body interaction that is smooth and slightly nonlocal. This does not affect the infrared properties of the model, and introduces an ultraviolet regularization at small scales, which is particularly convenient in the renormalization group analysis.

The Luttinger model plays a key role in the renormalization group analysis of the original lattice model (\ref{eq:H1d}). It captures the large-scale behavior of the correlation functions of the system, and its exact solvability allows to compute the critical exponents of the lattice model even in the absence of integrability; see \cite{Mabook} for a review of the results and of the renormalization group analysis. In particular, the Euclidean two-point function of the model (\ref{eq:H1d}) is given by, at zero temperature and in the infinite volume limit \cite{BMXYZ}:
\begin{equation}
\begin{split}
\big\langle {\bf T}\, a_{0} \gamma_{t} (a^{*}_{x}) \big\rangle_{\infty} &= g_{0}(t,x) \frac{1 + O(\lambda)}{(t^{2} + v^{2} x^{2})^{\frac{\eta}{2}}} + R(x,t) \\
g_{0}(t,x) &= \sum_{\omega = \pm} \frac{e^{i\omega k_{F} x}}{-it + \omega v x}\;,
\end{split}
\end{equation}
where: $v \equiv v(\lambda)$ is the interacting Fermi velocity, given at lowest order in $\lambda$ by the slope of the energy-dispersion relation at the Fermi level $\mu$; $R(x,t)$ is a faster-decaying error term as $t\to \infty$, $x\to \infty$; and $\eta = a_{0} \lambda^{2} + O(\lambda^{3})$ with $a_{0}>0$ is the anomalous exponent of the two-point function, analytic in $\lambda$ for $|\lambda|$ sufficiently small (the exact solution of the Luttinger model \cite{ML} provides the optimal existence and analyticity interval).

Renormalization group methods have been used to compute transport coefficients of $1d$ chains, in particular the susceptibility and the Drude weight \cite{BM1, BM2}. The starting point of the papers \cite{BM1, BM2} is the formulation of these transport coefficients in terms of Euclidean correlation functions, via the use of a Wick rotation, that can be rigorously justified \cite{MPdrude}. Let:
\begin{equation}
\begin{split}
D(\eta, p) &= \lim_{\beta \to \infty} \lim_{L\to \infty} D_{\beta, L}(\eta, p) \\ 
\kappa(\eta, p) &= \lim_{\beta \to \infty} \lim_{L\to \infty} \kappa_{\beta, L}(\eta, p)\;.
\end{split}
\end{equation}
Then, as proved in \cite{BM1, BM2, MPdrude, Mdrude2}:
\begin{equation}
\begin{split}
D(\eta, p) &= \frac{v K}{\pi} \frac{\eta^{2}}{\eta^{2} + v^{2} p^{2}} + R_{D}(\eta, p) \\
\kappa(\eta, p) &= \frac{K}{\pi v} \frac{v^{2} p^{2}}{\eta^{2} + v^{2} p^{2}} + R_{\kappa}(\eta, p)\;,
\end{split}
\end{equation}
where $K$ is the Luttinger parameter, related to the anomalous exponent of the two-point function by the formula $2\eta = K + K^{-1} - 2$, while the error terms $R_{D}, R_{\kappa}$ are continuous in a neighbourhood of $(0,0)$ and vanish in the limit $(\eta, p)\to (0,0)$. In particular, setting
\begin{equation}
D = \lim_{\eta \to 0^{+}} \lim_{p\to 0} D(\eta, p)\;,\quad \kappa = \lim_{p\to 0} \lim_{\eta \to 0^{+}}  \kappa(\eta, p)\;,
\end{equation}
we obtain the following remarkable identity between Drude weight, susceptibility and Fermi velocity:
\begin{equation}\label{eq:Hrel}
\frac{D}{\kappa} = v^{2}\;.
\end{equation}
The relation (\ref{eq:Hrel}) was first predicted to hold by Haldane in \cite{HL} on the basis of non-rigorous bosonization arguments, and it has been rigorously established in \cite{BFM1, BFM2, BM1, BM2}. The identity (\ref{eq:Hrel}) relates three non-universal quantities for non-integrable $1d$ systems in an exact way, and it is one of the defining properties of the Luttinger liquid universality class \cite{HL, HL2}. Recently, a similar `Haldane relation' has been discovered in the completely different setting of an interacting dimer model in classical statistical mechanics \cite{GMT20}, which turns out to belong to the Luttinger liquid universality class. In the context of classical statistical mechanics models, other remarkable identities for non-universal scaling exponents of non-solvable models belonging to the Luttinger liquid universality class have been proved in \cite{BFMexact}.

Finally, we conclude by mentioning that spinful fermions can also be studied, at the price of a considerably more involved renormalization group analysis \cite{BFM1, BFM2}. The main extra technical difficulty, solved in \cite{BFM1, BFM2}, is that the presence of the spin introduces further quartic marginal terms in the effective quantum field theory description of the model. These terms cannot be easily described in terms of emergent bosonic modes, and their renormalization group flow must be controlled via direct inspection of the beta function.

\subsection{Edge modes of interacting quantum Hall systems}

One-dimensional physics also arises at the boundary of two-dimensional topological insulators. The connection between the bulk topological order of insulating materials and the emergence of stable, metallic quasi-$1d$ modes is the content of the bulk-edge duality \cite{Hal}, a central concept in the theory of topological materials.

To begin, let us consider non-interacting quantum Hall systems, in the half-plane $\mathbb{Z} \times \mathbb{N}$. Let $H$ be the single-particle Hamiltonian, a self-adjoint operator on $\ell^{2}(\mathbb{Z} \times \mathbb{N}; \mathbb{C}^{M})$. We impose Dirichlet boundary conditions at $x_{2} = 0$. We shall also view the operator $H$ as the restriction to the half-plane of another Hamiltonian $H_{\text{B}}$, defined on $\mathbb{Z}^{2}$.

Let us suppose that $H_{\text{B}}$ is gapped, and that the Fermi level $\mu$ lies in a spectral gap. The bulk-edge duality establishes an identity between the value of the Hall conductivity of $H_{\text{B}}$, also called the bulk Hall conductivity, and the emergence of edge modes for $H$. For simplicity, let us further assume that both $H$ and $H_{\text{B}}$ are invariant under translations along the direction of the edge. Then, the Hamiltonians admit a partial Bloch decomposition,
\begin{equation}
H = \int^{\oplus}_{\mathbb{T}^{1}} \frac{d k}{2\pi}\, \hat H(k)\;,\qquad H_{\text{B}} = \int^{\oplus}_{\mathbb{T}^{1}} \frac{d k}{2\pi}\, \hat H_{\text{B}}(k)\;.
\end{equation}
Edge modes might appear for energies in the `bulk gap', corresponding to the spectral gap of $H_{\text{B}}$, and are associated to solutions of the Schroedinger equation for $H_{\text{B}}$, localized in the proximity of $x_{2} = 0$:
\begin{equation}\label{eq:edge}
\hat H_{\text{B}}(k) \varphi_{\omega}(k) = \varepsilon_{\omega}(k)  \varphi_{\omega}(k)\;,
\end{equation}
where $\varphi_{\omega}(k) \equiv \varphi_{\omega}(k, x_{2})$ is exponentially decreasing in the bulk,
\begin{equation}\label{eq:dec2}
| \varphi(k, x_{2}) | \leq Ce^{-cx_{2}}\;.
\end{equation}
The decay rate $c$ can be estimated in terms of the distance of $\varepsilon_{\omega}(k)$ to the rest of the spectrum of $\hat H(k)$. 

The edge modes become relevant for the zero-temperature transport properties of the system if the Fermi level intersects the edge modes dispersion relations $\varepsilon_{\omega}(k)$ at some value of the quasi-momentum $k$. We define the edge modes Fermi momenta as the solutions of:
\begin{equation}
\mu = \varepsilon_{\omega}(k_{F}^{\omega})\;.
\end{equation}
Around the Fermi level, the energy-momentum dispersion relation of the edge modes can be linearized, $\varepsilon_{\omega}(k' + k_{F}^{\omega}) - \mu = v_{\omega} k' + o(k')$, which suggests that the low energy excitations are effectively described by massless relativistic particles with velocities $v_{\omega}$, exponentially localized in proximity of the $x_{2} = 0$ edge, recall (\ref{eq:dec2}).

In this setting, the bulk-edge duality relates the value of the bulk Hall conductivity and the sum of the chiralities of the edge modes, defined as $\chi_{\omega} = v_{\omega} / |v_{\omega}|$, encoding the directions of propagation of the edge currents. This duality is encoded by the following amazing identity:
\begin{equation}\label{eq:be}
\sigma_{12} = \frac{1}{2\pi} \sum_{\omega} \chi_{\omega}\;.
\end{equation}
As we shall see below, the right-hand side of (\ref{eq:be}) can also be understood as an edge response function, which describes the variation of the edge current after introducing a perturbation localized at the boundary.

The first proof of (\ref{eq:be}) has been given in \cite{Hat}. It has then been extended to disordered systems whose Fermi energy lies in a spectral gap \cite{EG, SKR}, or in a mobility gap \cite{EGS}. More recently, the bulk-edge duality has been extended to other classes of topological insulators, such as time-reversal invariant systems \cite{GP}; see \cite{PS} for extensions to other topological phases. Field theoretic methods for the classification of topological phases have been introduced in \cite{F}, which allow to understand the bulk-edge duality in terms of anomaly cancellations between the Chern-Simons effective field theory for the bulk degrees of freedom, and the Luttinger liquid field theory arising on the boundary.

Let us now focus on rigorous results for many-body quantum systems. We shall consider weakly interacting fermions on the cylinder, namely on $\Gamma_{L} = [0, L]^{2} \cap \mathbb{Z}^{2}$ with periodic boundary conditions in the horizontal direction and Dirichlet boundary conditions at $x_{2} = 0,L$. We shall allow for internal degrees of freedom, and we shall denote by $\Lambda_{L}$ the decorated lattice; recall the discussion in Section \ref{sec:intro}. The points in $\Lambda_{L}$ are denoted by ${\bf x} = (x, \sigma)$, with $x\in \Gamma_{L}$ the space coordinate and $\sigma \in \{1, \ldots, M\}$ the color label. Let $H$ be a single-particle Hamiltonian, displaying edge modes, in the sense of exponentially localized solutions of (\ref{eq:edge}), at $x_{2} = 0$ and $x_{2} = L$. As usual, we shall denote by $\mathcal{H}_{0}$ the second quantization of the Hamiltonian, and we shall consider many-body Hamiltonians of the form $\mathcal{H} = \mathcal{H}_{0} + \lambda \mathcal{V}$, with $\mathcal{V}$ a short-ranged many-body interaction. We also suppose that the model is invariant under translations in the horizontal direction, which will allow us to perform a partial Bloch reduction.

We are interested in the edge transport properties of this model. Let $j_{i,{\bf x}}$ be the current density, such that $\mathcal{J}_{i} = \sum_{{\bf x} \in \Lambda_{L}} j_{i,{\bf x}}$, and satisfying the lattice continuity equation:
\begin{equation}
\partial_{t} \tau_{t}( n_{{\bf x}}) + \sum_{i=1}^{2} \text{d}_{i} \tau_{t} (j_{i,{\bf x}}) = 0\;,
\end{equation}
with $\text{d}_{i}f(x) = f(x) - f(x-e_{i})$ the discrete derivative in the $i$-th direction. We define the edge current $\mathcal{J}^{\ell}_{x_{1}}$ as:
\begin{equation}
\mathcal{J}^{\ell}_{x_{1}} := \sum_{\sigma} \sum_{x_{2} \leq \ell} j_{1,(x, \sigma)}\;.
\end{equation} 
This operator is associated with the charge current flowing in a strip of width $\ell \ll L$, adjacent to the $x_{2} = 0$ boundary. In order to probe the response of such current, we introduce the time-dependent Hamiltonian:
\begin{equation}
\mathcal{H}(\eta t) = \mathcal{H} + e^{\eta t} \sum_{{\bf y} \in \Lambda_{L}} \mu(y) a^{*}_{{\bf y}} a_{{\bf y}}
\end{equation}
where $\mu(y) \equiv \mu(y_{1})$ for $y_{2} \leq \ell'$  and $\mu(y) = 0$ otherwise. We will be interested in the variation of the average edge current, at first order in the external perturbation, in the range of parameters $1\ll \ell \ll \ell' \ll L$.

We define the edge response function as:
\begin{equation}
\begin{split}
G^{\ell, \ell'}_{\beta, L}(\eta, p) &:= -\frac{i}{L} \int_{-\infty}^{0} dt\, e^{\eta t} \Big\langle \Big[ \tau_{t}\Big(\hat n^{\ell'}_{-p}\Big), \hat{\mathcal{J}}^{\ell}_{p_{1}} \Big] \Big\rangle_{\beta, L} \\
G^{\ell, \ell'}(\eta, p) &:= \lim_{\beta \to \infty} \lim_{L\to \infty} G^{\ell, \ell'}(\eta, p)\;,
\end{split}
\end{equation}
where $\hat n^{\ell'}_{p} = \sum_{\sigma} \sum_{x_{2} \leq \ell'} \hat{n}_{p_{1}, x_{2}, \sigma}$ (the Fourier transform is over the variable $x_{1}$). In terms of this function, the linear response of the edge current can be written as:
\begin{equation}
\Tr\, \mathcal{J}^{\ell}_{0} \rho(0) = \frac{1}{L} \sum_{p} \hat \mu(p) G^{\ell,\ell'}_{\beta, L}(\eta, p) + O(\mu^{2})\;.
\end{equation}
We will be interested in the edge response function in the limits $\eta \to 0^{+}$, $p\to 0$ and then $\ell, \ell'\to \infty$ with $1\ll \ell \ll \ell' \ll L$, which is the relevant setting for studing slowly varying edge perturbations. Notice that the order of this limits matters: if one first takes $p\to 0$ and then $\eta \to 0^{+}$ one finds a trivial result: $\lim_{\ell'\to \infty} \lim_{p\to 0}G^{\ell, \ell'}(\eta, p) = 0$.

From a quantum field theory viewpoint, the large scale properties of the edge modes are expected to be described by a generalization of the Luttinger model, called the multi-channel Luttinger model, describing an arbitrary number of chiral relativistic fermions, interacting via a density-density type interaction. The Hamiltonian of the model is:
\begin{equation}
\begin{split}
&\mathcal{H}_{mL} = \sum_{\omega = 1}^{M} \int dx\, \overline{\psi}_{\omega, x} i c_{\omega} \partial_{x} \psi_{\omega, x} \\&\; + \sum_{\omega,\omega'} \lambda_{\omega, \omega'} \int dxdy\, \overline{\psi}_{\omega, x} \psi_{\omega, x} w(x-y) \overline{\psi}_{\omega', y} \psi_{\omega', y}\;,
\end{split}
\end{equation}
where the sum runs over the labels of $M$ chiral fermions, representing the edge modes at the Fermi level, with velocities $c_{\omega}$, and the couplings $\lambda_{\omega,\omega'}$ describe the edge mode scattering. As for the Luttinger model (\ref{eq:HL}), this quantum field theory can be studied via bosonization. It is the starting point of the chiral Luttinger liquid theory of edge modes \cite{Wen}, a non-rigorous field theory approach to edge transport in the quantum Hall effect.

Recently, the validity of this effective quantum field theory for the large-scale properties of edge currents has been rigorously proved; see \cite{AMP} for the case of one edge mode, \cite{MPspin} for two counterpropagating edge modes with opposite spins, and \cite{MPmulti} for the generic case of an arbitrary number of edge states. A key ingredient of the analysis is the vanishing of the beta function for the multi-channel Luttinger model \cite{MPmulti}, which allows to control the flow of the quartic marginal couplings, and to prove that the correlation functions decay with interaction-dependent anomalous exponents.

The renormalization group analysis, combined with Ward identities for the lattice model and for the effective quantum field theory, actually allows to compute the limiting edge response function. We obtain \cite{MPmulti}, for $|\lambda|$ small enough, and for $1\ll \ell \ll \ell'$:
\begin{equation}\label{eq:edgeuniv}
\lim_{p\to 0} \lim_{\eta \to 0^{+}} G^{\ell, \ell'}(\eta, p) = \sum_{\omega} \frac{\chi_{\omega}}{2\pi} + O(e^{-c\ell})\;,
\end{equation}
with $\{\chi_{\omega}\}$ the chiralities of the non-interacting edge modes.  In particular, taking the limits $\ell' \to \infty$ followed by $\ell\to \infty$, we see that the edge response function is quantized, and it is equal to the sum of the chiralities of the edge modes. Thus, combined with the universality of the Hall conductivity \cite{GMPhall}, Eq. (\ref{eq:edgeuniv}) allows to lift the bulk-edge duality to the realm of weakly interacting quantum Hall systems. Furthermore, the method also allows to compute the edge Drude weight and the edge susceptibility. In the case of one and two edge modes, this has been done in \cite{AMP, MPspin}. In particular, these results allow to check the validity of the Haldane relation (\ref{eq:Hrel}) for the edge response functions in these cases.

Other ways of probing edge currents are possible, involving different material geometries. For instance, the two-terminal conductance is obtained by connecting a Hall bar to two leads at different chemical potential, and by measuring the current propagating in the bar at first order in the difference of the chemical potential of the source and of the drain. For clean and non-interacting samples, it is actually expected that the two-terminal conductance is given by the sum of the absolute values of the chiralities. This quantity is not expected to be protected against many-body interactions between counterpropagating edge modes \cite{KF, KFP}. A possible mechanism to restore quantization in a way compatible with the bulk-edge duality relies on the introduction of disorder \cite{KF, KFP}, a relevant perturbation in the renormalization group sense. See \cite{Coh} and references therein for recent experimental studies of the two-terminal conductance in the context of the fractional quantum Hall effect. The rigorous understanding of this phenomenon is an interesting open problem in mathematical physics. 

\medskip

\noindent{\bf Acknowledgements.} A. G. and M. P. gratefully acknowledge financial support of the European Research Council (ERC) under the European Union's Horizon 2020 research and innovation programme (ERC CoG UniCoSM, grant agreement n. 724939 for A.G.; and ERC StG MaMBoQ, grant agreement n.80290 for M. P.). A.G. and V.M.  gratefully acknowledge financial support of MIUR, PRIN 2017 project MaQuMA, PRIN201719VMAST01. Our work has been carried out under the auspices of the GNFM of INdAM.

\end{document}